# Dynamic Simulation of Electro-Hydrodynamically Interacting and Sedimenting Particles

Sagardip Majumder, Jayabrata Dhar and Suman Chakraborty[1]
[1]Department of Mechanical Engineering, Indian Institute of Technology Kharagpur
Kharagpur 721302, INDIA

Particle-particle interactions in sedimenting systems have been investigated in the present study considering the many-body hydrodynamic and electrodynamic interactions. These interactions primarily occur in two modes: near-field and far-field interactions. The hydrodynamic interactions are modeled employing the Stokesian Dynamics while the electrodynamic interactions are accounted using the grand Capacitance matrix formulation capable of tackling externally applied arbitrary electric field effects. It is seen that the presence of an external electric field and asymmetry in particle positioning greatly modifies the dynamics of the rigid dielectric spherical particles when compared with the sedimenting system without the electric field effects. This is attributed to the induced dipole moment interactions among the particles. A consequence of the alterations in the particle arrangements also changes the net drag force experienced by these sedimenting particles, which is also reported in the present study. Furthermore, we have evaluated the induced background velocity field of the continuous medium due to sedimentation of the particles. A net velocity is observed in the continuous medium due to the sedimentation-induced particle rotation, which is found to vary in presence of an external electric field.

1. Introduction

Transport characteristics of finite numbers of particles sedimenting in the presence of a gravitational field and subjected to an external electric field as well as an imposed shear flow are of immense interest in the paradigm of low Reynolds number hydrodynamics. Electro-hydrodynamics of sedimenting particles holds utmost importance in many practical applications such as sedimentation of charged colloids (Raşa & Philipse 2004), particle sorting (Di Carlo et al. 2007), electrorheological suspensions (Bonnecaze & Brady 1992), sedimenting colloids including charged particles (Russell et al. 2012; Chang & Keh 2013), charged polymers and the study of their rheological behavior, porous medium transport, functionalities of protein solution (Brady & Bossis 1988; Mitchell & Spagnolie 2015), electromagnetokinetic transport of neutral particles (Kolin 1953), particle clustering and its control in sedimentation of colloids (Newman & Yethiraj 2015; Sullivan et al. 2003; Brayshaw et al. 1983) and controlled manipulation of sedimenting droplet under electric field (Xu & Homsy 2006), to name a few.

---

[1]Corresponding author, e-mail: suman@mech.iitkgp.ernet.in

In the literature, numerous studies have delineated the effects of many-body particle-particle hydrodynamic interactions on the resulting transport characteristics of particles through a viscous medium that is subjected to an external force (Durlofsky et al. 1987; Bossis & Brady 1984; Bonnecaze & Brady 1992; Tanaka & Araki 2000). A seminal work on sedimenting particle through a viscous fluid at low Reynolds number was reported by Stokes (Stokes n.d.). Subsequently, several studies reported extensions to this fundamental understanding. Faxén Law provided a generalized formula for the force and the torque on a spherical particle placed in an arbitrary background flow (Batchelor & Green 1972). Exact solution for the velocity field for two hydrodynamically interacting sedimenting spherical particles was subsequently obtained (Batchelor & Green 1972). Kynch (Kynch 1959) employed the method of reflection to obtain the effect of third and fourth interacting body. A Fourier-space multipole expansion was developed to address the sphere mobility functions for finite particle systems. Ladd (Ladd 1988) calculated the high-order multipole terms in the context of suspension and formed an approximate summation for certain many-body interactions. Later on, the spatial and temporal variations of the particle positions were addressed in view of a dynamical analysis, incorporating many-body interactions (Brady & Bossis 1988). The underlying fundamental understanding has been subsequently employed to understand the behavior in many practical scenarios, such as paper manufacturing (Steenberg & Johansson 1958), sedimentation of contaminant particles through an oil medium in internal combustion engines (Guazzelli 2006), and blood flow (Caro 2012). One critical aspect of such studies lies in comprehending and predicting the equilibrium macroscopic flow properties of sedimenting particles directly from its corresponding microscopic structure (Brady et al. 1988). It needs to be noted in this context that researchers have also comprehensively investigated the alterations in the flow properties in presence of an imposed shear (Durlofsky et al. 1987), or an arbitrary electric field for the case of colloidal electrorheological fluids (Bonnecaze & Brady 1992; Dhar et al. 2013). Such pertinent flow properties may be the hydrodynamic drag forces on particles (or particle chains) in the case of finite sedimenting particles or the suspension rheology (Durlofsky et al. 1987; Brady et al. 1988), sedimentation, or aggregation rate (Brady & Bossis 1988) and re-suspension electric field for the case of agglomerating colloidal suspension (Ramachandran 2007).

Bossis and Brady developed a comprehensive molecular dynamics-like simulation technique called the Stokesian Dynamics (Bossis & Brady 1984; Bossis & Brady 1987) that directly included the effect of short-ranged particle lubrication interactions, in addition to the long-range many-body interactions. This was obtained by ingeniously summing the inverse mobility matrix, comprising the first few moments of the particle multipole expansion, relating sphere velocities to forces and torques, eventually through the lubrication resistance matrix (Arp & Mason 1977; Jeffrey & Onishi 1984; Kim & Mifflin 1985), and subtracting the two body mobility matrix signifying far-field interactions. This results in the grand resistance matrix that relates the sphere forces and torques to its velocities.

Although reported studies have been concerned with many scenarios of particle dynamics with different particle shapes and sizes (Xia et al. 2009; Claeys & Brady 1993; Jamison et al.

2008), the effect of arbitrary electric field and imposed shear flow on sedimenting particles and the resulting flow in the background fluid matrix has not been explicitly investigated. Studying particle sedimentation, under such imposed conditions, may however, lead to a generalized electro-hydrodynamic description that may be extrapolated to the investigation of the dynamics of colloidal suspensions in a wide gamut of applications, ranging from the geophysical to the biomedical paradigm.

In the present study, we consider three-dimensional analysis of the sedimentation of finite sized solid particles in an infinite fluid domain in the low Reynolds number regime. The sedimenting particles may be subjected to a constant arbitrary electric field, or an imposed shear or vortex flow. The interaction of a sedimenting particle with either a neutrally buoyant particle or a wall particle (particle whose position remains fixed in the domain) is also studied under such flow conditions. When an electric field is applied in a system comprising finite numbers of particles immersed in a dielectric fluid medium, a particle-particle polarization force is induced due to the mismatch in the electrical conductivity of the particle and the continuous medium. By virtue of such interparticle electrostatic forces, the particles tend to form chain-like structures along the direction of the imposed field. However, the presence of other particles in the system influences the net force on the particle due to the enhanced local field. Extending the concept of Stokesian dynamics, a similar formulation, corresponding to the hydrodynamic interactions, to generate the many-body electrostatic interaction matrix among the particles, known as the Capacitance matrix, accordingly, is evaluated (Bonnecaze & Brady 1990). As an illustrative example, we provide a complete description of the details of the Capacitance matrix as a function of the position of the mobile and the wall particles subjected to an arbitrary constant or linearly varying electric field. Further, we obtain the velocity field of the continuous suspending medium due to particle motion under different flow scenarios. Before constructing the quantitative basis of the particle sedimentation study, we intend to qualitatively emphasize on the non-trivial nature of the problem wherein we must highlight the fact that the resultant particle transport due to coupled electrohydrodynamic forces is not a simple superposition of particle transport considering separately the hydrodynamic interactions and the electric field interactions. This coupled interconnection, despite an inherent linearity in the individual problems, stems from the non-linearity that arises due to the presence of many-body electrohydrodynamic interactions among the particles. In other words, the mathematical use of the position-dependent Resistance and Capacitance matrices denotes an intricate non-linearity to the resultant problem, and therefore, it cannot be viewed as a trivial extension of the hydrodynamic and electrodynamic particle transport mechanism. The consequent results may have far-reaching implications in understanding the mechanistic behavior of a system of particles subjected to an arbitrary electro-hydrodynamic field, in several applications of contemporary and emerging relevance.

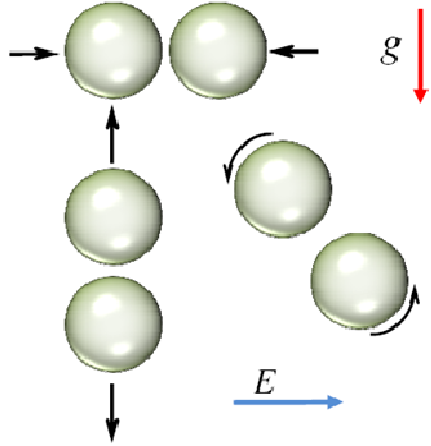

FIGURE 1. A schematic representation of the particle sedimentation under the effect of gravity (denoted by acceleration due to gravity g) phenomena in the presence of an external electric field. The electric field (*E)* induces a polarization effect due to mismatch in fluid and particle conductivities, which results in electrodynamic interactions among the particles. These interactions depend on particle orientation with respect to the electric field. The three possible major orientations are depicted in the figure. The other orientations are interpolations of these three major orientations in space.

## 2. Mathematical Methodology

We consider a three dimensional domain containing $N$ rigid spherical particles of radius $a$, density $\rho_p$ and permittivity ( conductivity) $\varepsilon_p$ ($\sigma_p$), suspended in a continuous medium having a viscosity $\eta$, density $\rho_f$ and permittivity (conductivity) $\varepsilon_p$ ($\sigma_p$). The particle size is assumed small enough so that the resulting Reynolds number $Ua/\nu$ becomes much less than unity, where $U$ is the characteristic particle velocity, $a$ is the characteristic size of the particles and $\nu = \eta/\rho_f$ is the kinematic viscosity. The particles are considered to be suspended in an unbounded Newtonian fluid subjected to an arbitrary electric field and a vertically downward gravitational field. The fluid domain may be stationary or undergoing an imposed uniform flow or linear shear flow. Here we intend to study the cumulative effect of hydrodynamic and electrostatic interactions among the sedimenting particles under gravity, observing the phenomenon of particle chaining and its influence on the effective drag force due to such chain formation. Various arrangements of such chains and agglomeration of particles due to an externally applied electric field may be found in situations related to electric field induced flocculation, electrorheological fluids and dielectrophoretic motion in a suspension (Halsey 2011; Bonnecaze & Brady 1992; Parthasarathy & Klingenberg 1996)

The simulation of particles accounts for both the many-body hydrodynamic interaction and the many-body electrodynamics. In the present study, we employ the Stokesian Dynamics method to simulate the particle hydrodynamics through the formulation of the resistance and the grand mobility matrices. For the electrodynamics, the formulation through the potential and the capacitance matrices incorporates the many-body interactions. In general, the many-body effects have two specific components that individually address the far-field and the near-field interactions among the particles. The dynamical evolution of the N-rigid particles is governed by the molecular dynamics-like force balance equation given by the Newton's second law of motion as

$$\mathbf{m}\frac{d\mathbf{U}}{dt} = \mathbf{F}^H + \mathbf{F}^G + \mathbf{F}^E + \mathbf{F}^{rep} \qquad (2.1)$$

Here $\mathbf{m}$ denotes the generalized mass (moment of inertia) matrix, $\mathbf{U}$ represents the particles translational (rotational) velocity vector and $\mathbf{F}$ describes the generalized force (torque) vector exerted on the particles. We consider four different forcing influences on the particles, namely, the hydrodynamic force $\mathbf{F}^H$ on the particles due to their motion relative to the fluid and other particles, the electrostatic force $\mathbf{F}^E$ due to the presence of an external electric field and fluid-particle dielectric/conductivity mismatch, the gravitational force $\mathbf{F}^G$ present uniformly throughout the domain, and the repulsive force $\mathbf{F}^{rep}$ that is introduced in the formulation to simulate rough particle surfaces and prevent particle overlaps in case of very high electric field strengths under a suitably chosen time step. This repulsive interaction is dominant within a threshold distance from a particle center. In this paper, we have neglected Brownian forces in the light that large particles are dominated by gravitation or electrostatic energies compared to the thermal energy within the system.

## 2.1. *Particles Electrostatic Interaction*

In order to simulate the electrostatic interparticle interactions accurately, the many body potential problem incorporating the Maxwell stress evaluation on each particle must be solved. However, such integration over particle surface is infeasible for more than two particles. Therefore, an alternative method to capture many-body interaction by using the concept of grand potential and capacitance matrix is pursued, which is simple yet highly accurate and implicitly takes into account the locally induced electric field and polarization effects among the particles. The method for evaluation of interparticle forces is achieved through the estimation of the electrostatic energy of the system, whose negative gradient with respect to the particles position generates the polarization force due to other particles in the system.

In the present study, we consider spherical charge free particles. The electrostatic energy density for *N* particles in a volume *V* is classically given by the form (Bonnecaze & Brady 1992)

$$\mathfrak{F} = \frac{1}{2V}\sum_{\alpha}\mathbf{S}_E^{\alpha}\cdot\mathbf{E} = \frac{1}{2V}\mathbf{S}_E\cdot\mathfrak{E} \qquad (2.2)$$

where $\mathbf{E} = E_0 \hat{\mathbf{E}}$ is the externally imposed electric field variable with $\hat{\mathbf{E}}$ being the unit vector for the electric field and $\mathbf{S}_E^\alpha$ is the induced dipole on particle $\alpha$. In the $N$-particle vector notation, $\mathbf{S}_E$ represents the 3N dipole vector of all the particles and $\mathfrak{E}$, the corresponding *3N* electric field vector. The relation of the particle dipole and the electric field can be obtained exploiting the linearity of the Laplace equation governing the potential distribution among the particles. Towards this, one can express a relationship between particle charge and dipole to the particle potentials and the gradient of the potential through a grand capacitance matrix given by (Bonnecaze & Brady 1991)

$$\begin{bmatrix} \mathbf{q} \\ \mathbf{S}_E \end{bmatrix} = \mathfrak{C} \begin{bmatrix} \Phi \\ \mathfrak{E} \end{bmatrix} \tag{2.3}$$

Here $\mathfrak{C} = \begin{pmatrix} \mathbf{C}_{q\phi} & \mathbf{C}_{qE} \\ \mathbf{C}_{S\phi} & \mathbf{C}_{SE} \end{pmatrix}$ is the $9N \times 9N$ capacitance matrix which not only depends on the particle positions but also is a function of the particle-to-fluid dielectric (conductivity) ratio. The capacitance matrix includes the many-body electrostatic interactions among the particles. Since we have considered charge free particles, we have $\mathbf{S}_E = \overline{C}(\mathbf{x}) \cdot \mathfrak{E}$ where the equivalent dipole electric field capacitance matrix is given by $\overline{\mathbf{C}}(\mathbf{x}) = \mathbf{C}_{S\phi} (\mathbf{C}_{q\phi})^{-1} \mathbf{C}_{\phi E} - \mathbf{C}_{SE}$ which is evaluated at instantaneous particle positions. The electrostatic energy density now has the equivalent form

$$\mathfrak{F} = \frac{1}{2V} \mathfrak{E} \cdot \overline{\mathbf{C}}(\mathbf{x}) \cdot \mathfrak{E} \tag{2.4}$$

The effective electrostatic force on particle $\alpha$ is, therefore, the gradient of the electrostatic energy as (Bonnecaze & Brady 1992)

$$\mathbf{F}^{\alpha E} = -V \frac{\partial \mathfrak{F}}{\partial \mathbf{x}_\alpha} = -\frac{1}{2} \mathfrak{E} \cdot \frac{\partial \overline{\mathbf{C}}(\mathbf{x})}{\partial \mathbf{x}_\alpha} \cdot \mathfrak{E} \tag{2.5}$$

where $\mathbf{x}_\alpha$ is the particle positions. For pairwise electrostatic particle forces, the energy is differentiated with respect to $\mathbf{x}_\alpha - \mathbf{x}_\beta$. It must be noted that the electrostatic force depends on the square of the electric field which is characteristic to induced polarization forces that gives rise to the dielectric electrorheological effect.

The aforementioned grand capacitance matrix must be so constructed that it incorporates the long-range many-body electrostatic interaction as well as short-range lubrication-like effects. The derivation of the capacitance matrix begins with the consideration of the integral form of the Laplace equation governing the potential distribution (Bonnecaze & Brady 1990; Bonnecaze & Brady 1991). The multipole expansion of the potential distribution, along with the expanded moments of the potential gradient and the gradient of the potential gradient, are evaluated. Combining the above expanded moments, which are truncated at the quadrupole level, with the Faxén-like law for potential distribution and its gradient and divergence of gradient of potential, the grand potential matrix relating the charge/dipole to potential/external electric field is developed. The grand potential matrix is a function of the instantaneous position of the particles and describes the far-field many-body interactions. The near-field interactions are significant for purely conducting particles. However, our assumption of dielectric sedimenting particles renders the near-field interactions negligible. The potential matrix is inverted to obtain the grand capacitance matrix $\mathfrak{C}$ relating the potential and externally applied field to the change and particle dipoles. Detailed method of construction of the grand capacitance matrix is provided in Appendix A.

## 2.2. Particle Hydrodynamic Interaction

In the regime of low Reynolds number hydrodynamics where the viscous forces dominate the inertial effects, Stokesian dynamics governs the description of particulate motion. The particle force, torque and stresslets are related to the translational velocity, rotational velocity and rate of strain by the grand resistance matrix $\mathscr{R}$ given as (Durlofsky et al. 1987)

$$\begin{bmatrix} \mathbf{F}^H \\ \mathbf{S}^H \end{bmatrix} = \mathscr{R} \begin{bmatrix} \mathbf{U} - \mathbf{U}^\infty \\ -\mathbf{E}^\infty \end{bmatrix} \tag{2.6}$$

Here for $N$ particles, $\mathbf{U}$ represents the three components of the velocity vector and the three components of the rotational velocity vector for each particle ($6N$ column matrix), and $\mathbf{U}^\infty$ is the corresponding imposed flow velocity in the domain evaluated at the particle center; while $\mathbf{E}^\infty$ is the rate of impressed strain tensor on the fluid (same for each particle which is symmetric and traceless by virtue of continuity; $6N$ column matrix). $\mathscr{R} = \begin{pmatrix} \mathbf{R}_{FU} & \mathbf{R}_{FE} \\ \mathbf{R}_{SU} & \mathbf{R}_{SE} \end{pmatrix}$ is a $11N \times 11N$ grand resistance matrix. $\mathbf{F}^H$ Includes the hydrodynamic forces and torques for $N$ particles while $\mathbf{S}^H$ are the complementary stresslets on each particle.

In the above scenario, the hydrodynamic force that a particle experiences due to a bulk shear flow and the presence of other particles has the form

$$\mathbf{F}^H = -\mathbf{R}_{FU} \cdot (\mathbf{U} - \mathbf{U}^\infty) + \mathbf{R}_{FE} \cdot \mathbf{E}^\infty \tag{2.7}$$

The resistance matrices $\mathbf{R}_{FU}$ and $\mathbf{R}_{FE}$ are representatives of tensors relating the hydrodynamic forces (torques) on the particles to their relative motion with respect to the continuous medium and to the imposed shear flow, respectively. The resistance matrices depend only on the instantaneous particle position and geometry of the domain, whose evaluation may be found elsewhere (Durlofsky et al. 1987). Here we will outline the basic procedure of finding the hydrodynamic grand mobility and resistance matrices.

The integral representation of the velocity field at any point in the fluid or inside the rigid particles in a Stokes flow is (Kim & Karrila 1991)

$$u_i(\mathbf{x}) = u_i^\infty(\mathbf{x}) - \frac{1}{8\pi\eta} \sum_{\alpha=1}^{N} \int_{S_\alpha} J_{ij}(\mathbf{x} - \mathbf{y}) f_j(\mathbf{y}) dS_y \tag{2.8}$$

where $u_i^\infty(\mathbf{x})$ is the background velocity field in the absence of particles, $S_\alpha$ is the surface of particle $\alpha$ (where we consider there are $N$ particles) with $\mathbf{y}$ as the position on the particle surface, $\mathbf{x}$ being any field point on the whole domain. $J_{ij}$ is the Stokeslet or Oseen tensor given by $J_{ij}(\mathbf{r}) = \dfrac{\delta_{ij}}{r} - \dfrac{r_i r_j}{r^3}$ with $\mathbf{r} = \mathbf{x} - \mathbf{y}$ and $r = |\mathbf{r}|$. The force density on particle surface due to the fluid stress $(\boldsymbol{\sigma})$ is represented as $f_j(y) = \sigma_{jk}(y) n_k(y)$, with $\mathbf{n}$ being the surface normal vector. The integral equation of the velocity field is then expanded in moments about the particle centre where all the moments are required to incorporate the near-field lubrication effects besides the far-field hydrodynamic interaction. However, moments till the dipole term are taken in addition to two higher multipole contributions that incorporate the finite size effects of the particles to form the mobility matrix. The resulting velocity field at any point results in the expression (Durlofsky et al. 1987)

$$u_i(\boldsymbol{x}) = u_i^\infty(\boldsymbol{x}) - \frac{1}{8\pi\eta} \sum_{\alpha=1}^{N} \left( \left(1 + \frac{1}{6} a^2 \nabla^2\right) J_{ij} F_j^\alpha + R'_{ij} L_j^\alpha + \left(1 + \frac{1}{16} a^2 \nabla^2\right) K_{ijk} S_{jk}^\alpha \right) \qquad (2.9)$$

Here $R'_{ij} = \varepsilon_{ijk} \dfrac{r_k}{\mathbf{r}^3}$ is the torque propagator (rotlet), $F_j^\alpha$ and $L_j^\alpha$ are, respectively, the force and torque exerted by particle $\alpha$ on the fluid measured relative to particle centre. $S_{ij}^\alpha$ is the stresslet on particle $\alpha$ while $K_{ijk} = \dfrac{1}{2}(\nabla_k J_{ij} + \nabla_j J_{ik})$.

Combining the above procedure with the Faxén Law, the motion of any particle in the flow field can be evaluated by constructing a grand mobility matrix that accounts for the far-field particle-particle interactions and size effects, and relates the particle force/torque and stresslets to its velocity/angular velocity/rate of strain. The lubrication effects, which have been neglected due to truncation of the moments expansion, are included by addition of the two-sphere lubrication resistance matrix (Arp & Mason 1977; Jeffrey & Onishi 1984; Kim & Mifflin 1985) to the inverted mobility matrix and subtracting the far-field effects (the inverted two-body mobility matrix). The resulting grand resistance matrix $\mathscr{R}$ (as used in equation (2.6)) (Durlofsky et al. 1987) relates particle velocity/angular velocity/rate of strain to its force/torque and stresslets. The detailed description of the construction of the grand resistance matrix incorporating force-torque-stresslet (F-T-S) relation is provided in Appendix B. This method of analysis approximates the higher order moments and preserves the near-field lubrication effects. We shall employ equation (2.9) to evaluate the velocity field in the fluid domain and discuss the effect of incorporation of particle torques on the net particle displacement (and hence a net fluid flow) in the sedimentation process.

*2.3.    Immobile particle formulation*

Another aspect that has been included in the present formulation is the effect on mobile particles due to the presence of wall/immobile particles. The wall/immobile particles are initiated by specifying their position manually and fixing their velocities to zero. The same approach was used by Nott and Brady (Nott & Brady 1994) to simulate particle hydrodynamics in pressure driven flow by considering the walls to be clusters of closely packed stationary spherical particles. In case of mobile particles, the net force, torque and stresslet are given as inputs in each of the iterations in order to update the velocities. However, in the case of immobile particles, their velocities are specified as inputs and the net reactions on these particles (force, torque, stresslet) are calculated.

*2.4. Repulsive force interaction*

It has been mentioned in previous works (Klingenberg et al. 1989; Phung et al. 1996) that, although lubrication effects significantly reduce the radial component of particle velocities near contact, the true simulation of hard sphere interactions is incomplete without the presence of short-ranged mutual repulsive interactions. The rationale behind the inclusion of an heuristic formulation stands on the fact that such form for force estimation closely resembles the hard-sphere repulsion interaction with a characteristic rapid decay besides retaining the physical size of the rigid particles and simulate hard sphere particle-particle repulsive interactions with a pre-defined cut-off radius (Klingenberg et al. 1989). Further, similar forms have been employed in numerous studies in the literature (Nott & Brady 1994; Morris & Brady 1998; Klingenberg et al. 1989). For our purpose, we have included a heuristic exponentially decaying function that corresponds to the short-range core-core repulsive interactions of the Buckingham potential (Buckingham 1938; Lane et al. 2009; Kendall et al. 2004) for simulating such repulsive interactions: $\tilde{\mathbf{F}}_i^{\mathbf{rep}} = \sum_{j=1}^{N} \alpha \mathbf{e_{ij}} \exp\left(-100\left(r_{ij} - 2R\right)\right)$ (Klingenberg et al. 1989; Parthasarathy & Klingenberg 1999), where $\tilde{\mathbf{F}}_i^{\mathbf{rep}}$ represents the non-dimensional repulsive force on the $i^{th}$ particle due to interaction with all the other particles, $N$ being the total number of particles in the fluid, $\mathbf{e_{ij}}$ represents the unit vector denoting the position of the center of the $j^{th}$ particle with respect to the $i^{th}$ particle with $r_{ij}$ as their central distance, and $R$ denotes the radius of each spherical particle. As mentioned before, such an imposed force is essential in the present study in order to prevent particle overlaps when the charge polarization effects are quite dominant. The presence of this force helps in reducing the computation time by enabling us to choose a suitable time step that is larger than the one which would be required if $\mathbf{F}^{\mathbf{rep}}$ is not taken into consideration. The order of magnitude of this heuristic force is taken to be the same as that of the polarization force so that at the threshold distance from a particle center, the magnitudes of both these forces become identical. The parameter $\alpha$ determines the strength of the repulsive interaction which has been taken as unity in conjunction with previous works on electrorheological fluids (Klingenberg & Zukoski 1990; Klingenberg et al. 1991; Parthasarathy & Klingenberg 1996).

## 2.5. Gravitational force on Particles

The gravitational force is a volume force over the whole domain and is independent of the particle arrangements. For each particle, the effective gravitational force on it, taking into account the buoyancy effects, is given by

$$\mathbf{F}^G = \frac{4}{3}\pi a^3 \left(\rho_p - \rho_f\right) g \qquad (2.10)$$

The terminal velocity for a particle, when no other particle is present in the system, is given by the balance between the Stokes drag and the gravitational force on it. Later, it will be exploited to define the time scale for the sedimenting problem. For the case of neutrally buoyant particles, one may prescribe a net zero gravitational force on those respective particles, which will render them inert to gravitational effects. However, an interesting aspect of the present formulation will nevertheless impose a hydrodynamic and electric force on these particles which manifest in a vertical movement of these neutrally buoyant particles.

## 2.6. Particle Sedimentation

Ignoring the inertial effects for low Reynolds number flow limit, we neglect the left hand side of equation (2.1). The particle positions are updated by replacing equation (2.7) in equation (2.1), thereby obtaining a form of the coupled equation of motion for the sedimenting particles as

$$\frac{d\overline{\mathbf{x}}}{d\overline{t}} = \overline{\mathbf{U}} = \overline{\mathbf{U}}^\infty + \overline{\mathbf{R}}_{\mathbf{FU}}^{-1} \bullet \left(\left(Mn^{-1}\right)\left(\overline{\mathbf{F}}^{\mathbf{E}} + \overline{\mathbf{F}}^{rep}\right) + \xi \overline{\mathbf{F}}^{\mathbf{G}} + \overline{\mathbf{R}}_{\mathbf{FE}} \bullet \overline{\mathbf{E}}^\infty\right) \qquad (2.11)$$

Equation (2.11) is expressed in a dimensionless form where all the lengths is nondimensionalized by the characteristic particle radius $a$, the viscous forces, the effective gravitational forces and the electrostatic forces by $6\pi\eta a^2/t_0$, $\frac{4}{3}\pi a^3 \left(\rho_p - \rho_f\right) g$ and $12\pi\varepsilon a^2 \left(\beta E_0\right)^2$, respectively, where the time scale $t_0 \sim 9\eta / 2g\left(\rho_p - \rho_f\right)a$ is described as the time required for the particle to travel one radial distance at steady terminal velocity condition. $g$ is the acceleration due to gravity and $\varepsilon = \varepsilon_p \varepsilon_0$ with $\varepsilon_0$ is the permittivity of free space. Here $Mn = \dfrac{\eta/t_0}{2\varepsilon\left(\beta E_0\right)^2}$ is the Mason number signifying the relative importance of the viscous shear force to the electrostatic force and $\xi = \dfrac{2}{9}\dfrac{a\Delta\rho g t_0}{\eta}$ represents the ratio of the gravitational force to the viscous force on a particle. Since the simulation is performed under the steady state assumption, wherein the gravitational forces are balanced by the viscous forces for each particle, the value of $\xi$ is chosen as unity throughout the present work. In the remaining part of the paper, we refer to physical quantities in terms of the dimensionless variables; therefore, we drop the bar on the physical values for convenience.

Another interesting aspect of particle sedimentation is the drag force experienced by the particles, which deviates from the classical Stokes drag $6\pi\eta aU$ where $U = |\mathbf{U}|$. The comparison of this drag force is made through the estimation of drag coefficient $\lambda = F_D/6\pi\eta aU$ where $F_D$ is the evaluated drag force on the particles employing the Stokesian dynamics. With the present non-dimensionalization procedure, the drag coefficient may be recast as $\lambda = 1/U$ where $U$ is the dimensionless velocity of the particles.

3. Simulation Method

For simulating the sedimenting spherical particles invoking Stokesian Dynamics and the Dielectric Electrostatic interactions, we have used MATLAB 2014 as our programming platform. Unlike the Stokesian Dynamics simulation, the mobility and potential matrices were calculated at each time step for better accuracy and the integration of equation (2.11) was carried out using the 'Euler' numerical integration scheme (Durlofsky et al. 1987). We have employed a maximum dimensionless time step of 0.1. The value of the time increment $dt$ for different simulations was so chosen that the magnitude of the velocity $U$ remained of the order of one, for all times. This is essential in preventing large changes in the particle positions when their distance of separation is small due to the increased electrostatic forces near particle surfaces (Phillips 1996). The repulsive force on a particle was calculated considering a cut-off distance of 2.5 times the particle radius. Also, the ratio of the conductivities for the particles and the fluid medium was taken as 4 throughout the range of simulations, merely for illustration.

4. Results and discussion

The three-dimensional dynamic evolution of the finite sedimenting spherical particles and their arrangement in conditions of varied viscosity and electric field is explicated in the present section. We have considered different scenarios with 3 and 5 particles arranged symmetrically and asymmetrically in an infinite flow domain. The present formulation can take into account the effect(s) of an imposed uniform flow, or an arbitrary imposed shear, or a constant arbitrarily directed external electric field or any combination of the above. An interesting feature of the present formulation is that it also captures the interaction of a sedimenting particle with a neutrally buoyant particle or a static wall particle and the corresponding unsteady velocity field in the fluid. The details of the velocity field formulation from equation (2.9) have been provided in Appendix C. For the simulation of the sedimenting particles, we have varied the $Mn$ and $\xi$ to observe the contrasting particle trajectories and obtain the particle drag coefficient.

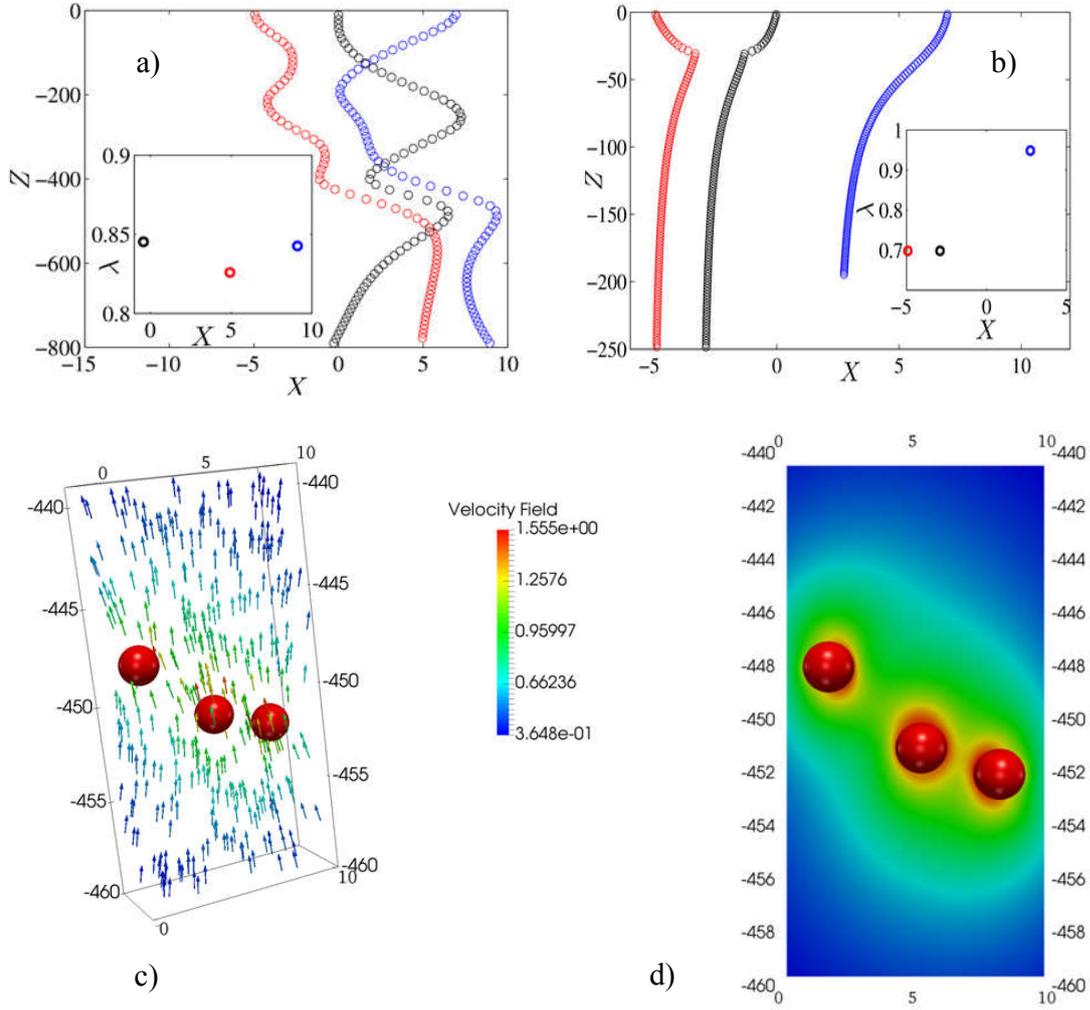

FIGURE 2: The sedimentation trajectory of three spherical particles under the influence of a) gravity and in absence of any external electric field, b) gravity and an externally applied field in the x-direction. The dimensionless parameters considered are a) $\xi = 1$; b) $\xi = 1$ and $Mn = 0.1$. Inset of the figures denote the drag coefficient for the respective sedimenting spherical particles. c) A 3-D vector view of the velocity field in the vicinity of the three particles corresponding to a) at a particular sedimenting position. d) A plot of the two dimensional velocity field for figure 2c in the plane y=0.

### 4.1. Asymmetrically-placed Particle Positions

Here we consider the case of three spherical particles, placed at an initial dimensionless position (-5,0,0), (0,0,0) and (7,0,0) and five spherical particles placed at initial dimensionless positions (-8,0,0), (-3,0,0), (0,0,0), (5,0,0) and (8,0,0), sedimenting under the effects of gravity and an externally applied electric field. We simulate the particle trajectories for the sedimentation phenomena with and without the application of an external electric field and evaluate the corresponding drag coefficient $(\lambda)$ for the particle clusters so formed.

Figure 2 denotes the trajectories for three spherical asymmetrically-placed particles sedimenting under the effect of the gravitational field in an infinite fluid domain. The figures are plotted for the following values of the dimensional parameters: a) $\xi = 1$ and $Mn^{-1} = 0$, signifying the case of no externally applied electric field; and b) $\xi = 1$ and $Mn = 0.1$, signifying the application of an external electric field of constant magnitude along the positive x-direction. The particle trajectories, for mildly altered configurations, show a sharp contrasting variation especially in the latter half of the sedimenting process where the effect of particle rotations dominates. Figure 2a denotes the trajectories of three sedimenting particles reproducing the result from the work of Durlofsky et. al (Durlofsky et al. 1987), when no external electric field is applied. The corresponding velocity arrow diagram, depicted in figure 2c, depicts the 3D velocity vector field of the fluid medium in the vicinity of the sedimenting particles, at a particular instant during the sedimentation process. The 2D velocity field of the fluid, at the same instantaneous position as in figure 2c, is depicted in figure 2d. It is seen that the velocity is the highest around the particles and gradually decreases as one move away from the particle centers. Similar flow trends around single sedimenting particle have been shown in previous studies (Drescher et al. 2010; Pozrikidis 2011). We shall use this figure to qualitatively compare the flow velocity of the fluid around the particles when an external electric field is imposed. As the electric field comes in the picture, a drastic change in the particle trajectories and resulting drag forces occur. The particles at the onset of the sedimentation tend to form clusters or chains along the electric field due to the induced polarization forces attributed to the mismatch of the particle and fluid conductivities. Figure 2b clearly depicts the particle chaining in the direction of an external electric field. The two left particles depicted early chaining along the field due to their close proximity as compared to the rightmost particle.

Insets of figures 2a and 2b describe the corresponding drag coefficients for the three spherical particles in the respective figures. Such asymmetric sedimentation trajectories stem from the initial particle positions. Three significant observations may be made from figure 2. First, from the linearity analysis of Stokes flow, one can rule out the phenomena of Magnus effect which occurs due to presence of particle translation and rotation in high Reynolds number flows (Leal 2007). However, here we find the particles tend to move in the direction orthogonal to the plane of the particle translation and the particle rotation vector even in the absence of an electric field. This may be attributed to the non-linearity introduced while accounting for the effect of many-body hydrodynamics due to the presence of other interacting particles (Leal 2007). Presence of an external electric field further augments the rate of transverse particle movement. Secondly, due to this non-linearity, there is a net displacement of the center of mass of the particle system which, in turn, manifests in a net flow rate of the surrounding fluid. Finally, as a consequence of these relative particle positioning, the hydrodynamic drag on each particle deviates from the classical Stokes drag for a particle in an unbounded fluid. As noted in previous studies, a particle sandwiched between other particles experiences less drag, which is apparent from the drag coefficient magnitude of the middle particle in Figure 2a. The other two particles experience higher drags due to the absence of particles on one of their sides. Their drag coefficients are close to unity (Stokes drag value) with reduced sedimentation velocities. However, with the application of an electric field, we find the particles form chains; thus, all the particles in the chain (here two particles) experience a similar drag, which is significantly less than that in the case without the electric field, and consequently the particles sediment with a higher velocity.

Figure 3a depicts the sedimentation of five asymmetrically placed particles subjected to an x-direction electric field with $\xi = 1$ and $Mn = 0.1$. It is again found that the four nearest particles quickly form a chain and start descending as a single entity while the left out particle experiences a larger drag, and thereby, possesses a lower velocity. It must be appreciated that the drags on each of the four clustered particles will be close enough in magnitude, but considerably less than the single sedimenting particle whose drag coefficient shall be close to unity. This is consistent with the plot shown in the inset of figure 2b. Figures 3c and 3d describe the velocity field in the vicinity of the four sedimenting particles. It can be deduced that the velocity of the surrounding fluid is maximum near the particles and reduces as one move away from the cluster. Further, the vector arrow plot for the fluid velocity qualitatively depicts the motion of the fluid domain around the particles. In fact, the four clustered particles must sediment with a higher velocity when compared to the case where electric field is absent. This can be seen qualitatively when compared with figure 2d (for three sedimenting particles in absence of electric field) that the maximum velocity is higher and the drag coefficients are lower in the present case with electrostatically-induced clustered particles.

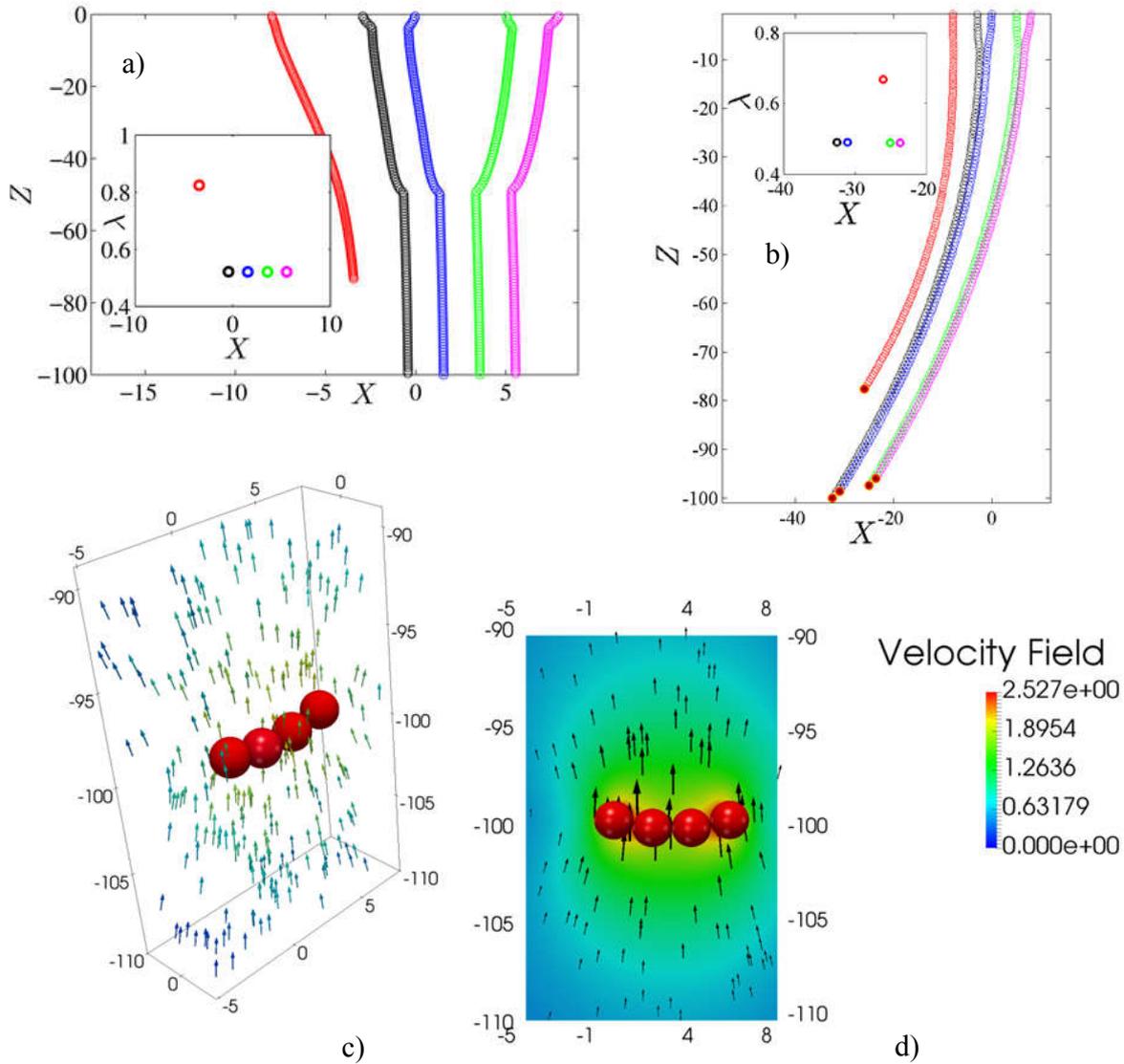

FIGURE 3: The sedimentation trajectory of five spherical particles under the influence of gravity and an externally applied field in the x-direction a) with no shear; and b) with an imposed shear of $\gamma = 0.01$. The dimensionless parameters considered are $\xi = 1$ and $Mn = 0.1$. Insets of the figures denote the corresponding drag coefficients for the respective sedimenting spherical particles. c) A 3-D vector view of the velocity field in the vicinity of the five sedimenting particles corresponding to a) at a particular sedimenting position where the chain has been developed. d) A plot of the two dimensional velocity field corresponding to c) in the plane y=0.

Figure 3b finally introduces the effect of an externally imposed shear on the sedimenting particles. In the present case, we have imposed a non-dimensionalized shear in the negative X-Z plane with $\gamma = 0.01$. With an imposed shear, the particles no more sediment in clusters but tend to form smaller two-particle chains at an early stage of the sedimentation process. These chains keep migrating along with the shear, maintaining the direction of chaining along the applied electric field, as shown. The inset of figure 3b shows the particle drags wherein the drags on the smaller chains are less than the drag experienced by the left out particle.

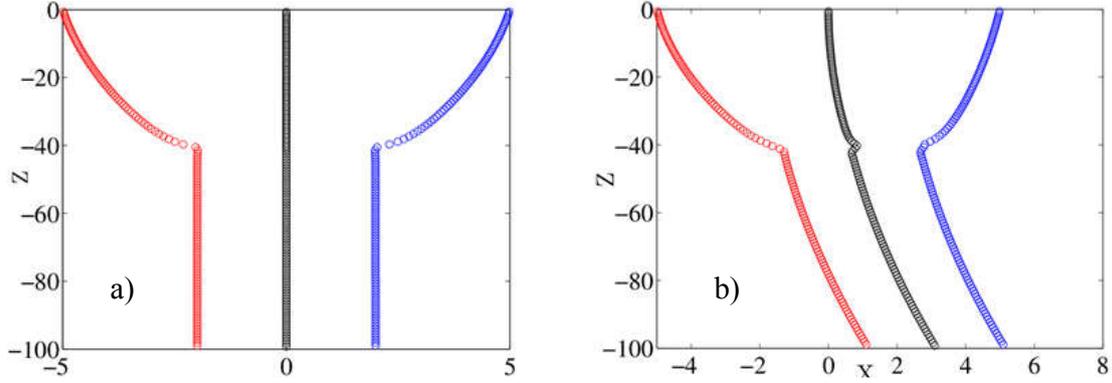

FIGURE 4: Plots representing the particle sedimentation of three particle system placed symmetrically in presence of an external electric field in a) X-direction; b) X-direction with a dimensionless shear of $\gamma = -0.001$ in the X-Z plane. The dimensionless parameters are $Mn = 0.1$ and $\xi = 1$.

### 4.2. Symmetrically-placed Particle Positions

Here we consider the case of three particles at an initial symmetrical dimensionless position (-5,0,800), (0,0,800) and (5,0,800). We study the particle trajectory and the drag coefficient $(\lambda)$ for different values of the dimensionless parameters, direction of electric field and imposed shear.

Figure 4 denotes the sedimentation trajectory of three particle system placed symmetrically in the fluid domain. Figure 4a classically shows the chain formation of the three particles in the direction of the electric field. An intriguing feature is that the formulation captures the symmetry of the system which is maintained even with the horizontally applied field. The drag of the particles, which will be discussed later, will thus be actually much less compared to the no electric field case due to chain formation (here drag coefficient $\lambda \sim 0.59$). Figure 4b depicts the particle trajectories in presence of a shear $\gamma = -0.001$ and an inclined electric field. The shear acts as a catalyst for the particle to form the symmetry and aligns them in the inclined electric field direction. Further, it can also be seen that due to the shear, the particles, although do not break their chains, align in a direction which is slightly offset from the actual 45 degrees inclination. This chain is strong enough to resist breakage even in presence of viscous, shear or gravitational effects, which inevitably gives rise to the electrorheological behavior of particle suspension in the presence of an electric field.

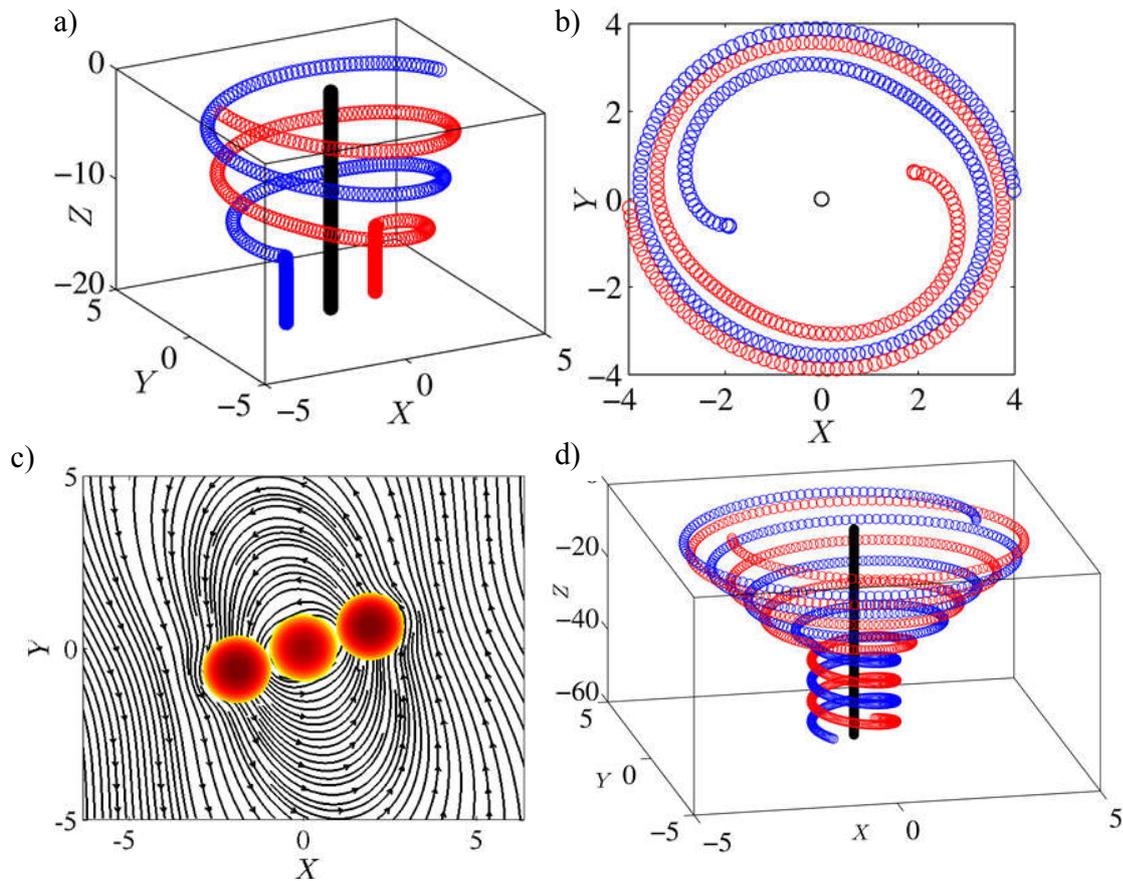

FIGURE 5: a)[b)] The isometric view [top view] of the trajectories of three sedimenting particle system placed symmetrically ([-4,0,0], [0,0,0], [4,0,0]) in presence of an imposed vortex flow with strength $\Omega = 1$ about the center particle and an external electric field in the X-direction with $Mn = 0.025$. c) The streamlines, corresponding to a), on the X-Y plane (at Z = -20) after particle revolution about the central particle has ceased. d) the isometric view of the trajectory of three sedimenting particle system placed symmetrically ([-3,0,0], [0,0,0], [3,0,0]) in presence of an imposed vortex flow with strength $\Omega = 1$ about the center particle and an external electric field in the Z-direction with $Mn = 0.01$.

### 4.3. Vortex Flows

In this section we examine the sedimentation of three particles under the condition of an imposed vortex flow. The trajectory of the falling particles shows interesting patterns which vary depending on the electric field magnitude, direction and the initial position of the particles. This study holds implications as to how particles clustering particles would behave in circulating flows.

Figure 5a and 5b show the isometric view and top view, respectively, of the trajectory of three symmetrically placed sedimenting particles in presence of an imposed vortex flow of strength $\Omega = 1$ and an external electric field with $Mn = 0.025$. The sedimenting particles follow a helical trajectory which is attributable to the coupled gravitational attraction towards the negative Z-axis and a polarization-induced attraction towards the center particle. The intriguing aspect of the trajectory is brought out when the particles approach each other and form a chain. Intuitively, one might expect that the particle chain must continue to revolve about the central particle. However, an interesting observation, as clearly seen from figure 5b, is that the particle revolution ceases after the chain is formed. The chain sediments vertically, making a specific angle with the X-Z plane. The particular orientation of the chain is due to the dynamic balance between the electrostatic torque the chain experiences, since the chain axis (the line joining the particle centers in the X-Y plane) makes a specific angle with the applied external field, and the viscous torque acting on the particle due to the applied vortex flow. As seen in figure 5c, the streamlines depict the direction of the flow that tends to orient the chain axis further away from the X-directed electric field. It must be noted that this angle the particle chain makes with the X-Z plane will vary with the Mason number. Low values of the Mason number suggest larger polarization effects, consequently a small angle. On the contrary, for higher $Mn$, the induced electric polarization torque is not strong enough to overcome the viscous torque and the particle chain exhibits a revolving motion about the central particle. In this regime, the viscous torque is always larger than the maximum electrodynamic torque, which occurs when the chain axis is perpendicular to the applied field direction. It is further noted that the streamlines cannot make a perfect circular vortex pattern about the central particle due to the presence of the two other particles; rather an elliptic streamline pattern is observed in the vicinity of the particle chain. Figure 5d depicts an interesting case when the electric field is applied in the Z-direction instead. Even at such low $Mn$ value, the particle chain maintains the revolving motion which is attributable to the fact that the field is directed perpendicular to the plane of viscous torque. In fact, a closer look will reveal that the particles in this case re-orient to form an inverted triangular cluster which revolves about the center particle.

### 4.4. *Some General Discussions*

Figure 6 discusses about three different cases for sedimenting particles. Figure 6a shows the case of three asymmetrically placed spherical particle sedimenting under a gravitational field and subjected to a vertical electric field (in the z-direction) with $Mn = 0.001$. Due to strong electrostatic repulsive forces, the particles initially move away from each other; but later on the particles eventually rotate and tend to align themselves along the direction of the electric field. This shows a very interesting case of variation of particle motion and alignment due to the coupled hydrodynamic and electrostatic effect. Figure 6b depicts the similar case of symmetrically-positioned particles sedimentation but with an electric field in the X-Z direction with $Mn = 0.001$. Due to such a strong electrostatic force, it is intriguing to see how the rightmost particle actually moves up owing to the strong attraction along the direction of the applied electric field while the leftmost particle is dragged down faster. Eventually, the three particles form a chain along the applied electric field direction and sediment downwards maintaining that structure. It must be reiterated that the chaining direction depends on the relative dominance of the gravitational and polarization forces. Stronger gravitational force would imply that the angle the particle chain makes with the horizontal deviates from $45^0$ while a stronger polarization forces means the chain aligns closely with the applied electric field that is directed at $45^0$ with the horizontal in the present case.

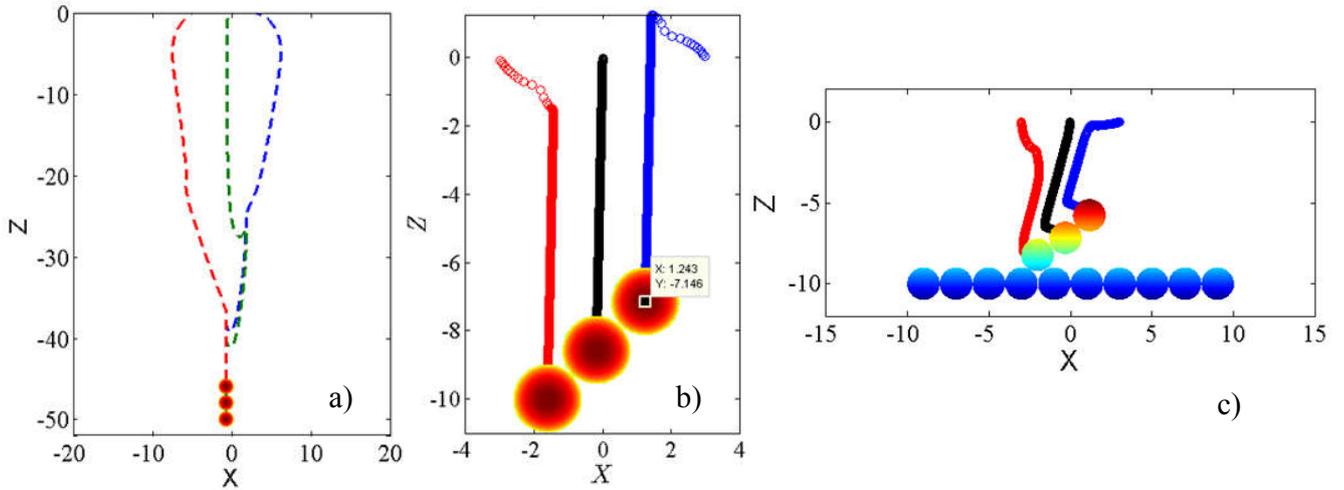

FIGURE 6: a) Plot of the particle trajectories for three sedimenting particles in the presence of an external field in Z-direction with $Mn = 0.001$. The Initial positions of the particles were (-5,0,0), (0,0,0) and (3,0,0). b) Plot of the particle trajectories for three sedimenting particles in the presence of an external field in X-Z-direction with $Mn = 0.001$. The Initial positions of the particles were (-3,0,0), (0,0,0) and (3,0,0). c) Plot of the particle trajectories for three sedimenting particles in the presence of a horizontal string of immobile particles with their centers at $z = -10$ downstream and an external field in X-Z-direction with $Mn = 0.01$. The Initial positions of the particles were (-3,0,0), (0,0,0) and (3,0,0).

Figure 6c shows an interesting case of the interaction of mobile particles with rigid immobile particle (or particle string). The figure depicts a 3D view of the particle trajectories when a string of immovable particles is introduced in their sedimenting path. The mobile particles form a chain initially owing to the strong electric field, and the chain finally attaches into one of the grooves between two immobile particles. The formulation captures the intricate upstream effect of the string of immobile particles due to which the particle trajectories actually deviate when compared to figure 6b that has no such particle strings placed downstream.

The observations made for all the above arrangements may be interestingly related to the electrorheological characteristics that particulate suspensions exhibit, in presence of an external electric field and a shear induced flow. Increasing the number of mobile particles and/or applying an electric field in other directions will explore other various flow arrangements, which, for the sake of brevity, have been excluded from the present analysis. With this view that different magnitudes and direction of the imposed electric field drastically alter the sedimentation characteristics of the particles, we proceed to discuss the variation of the drag coefficient and the net fluid displacement rate as a function of the Mason number for electric field applied in three different directions. The fluid displacement rate hitherto referred to as the volume flow rate, is a qualitative measure of the instantaneous rate of flow of the displaced fluid in the horizontal direction (X-axis) opposite to that corresponding to the movement of asymmetrically positioned particles.

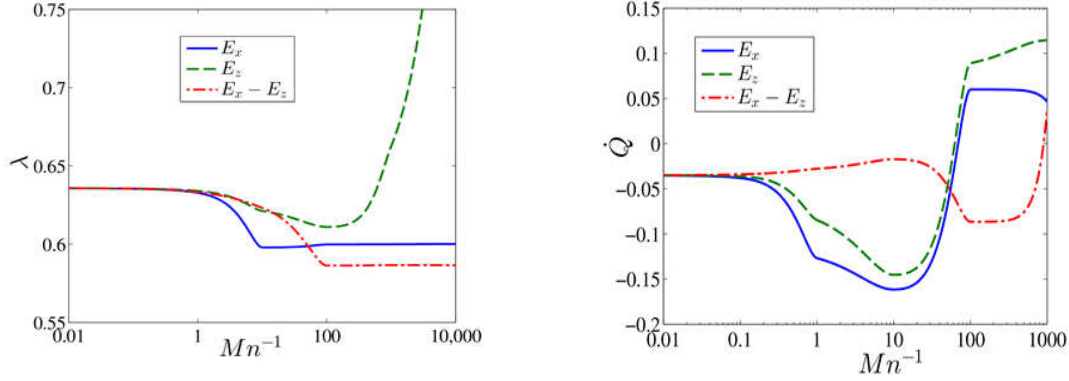

FIGURE 7: Plots representing the variation of the a) drag coefficient $\lambda$, and b) volume flow rate $\dot{Q}$ as a function of the Mason number for different orientations of the applied electric field.

Figure 7a denotes the variation of the drag coefficient at a particular vertical position $(z=10)$ of the middle particle in a symmetrically placed three particle system, as a function of the inverse of Mason number for three different electric field orientations. It is seen for lower $Mn^{-1}$ values, signifying lower domination of the polarization effects, that the direction of electric field has no effect on the drag coefficient. In this region, the drag coefficient of the middle particle is similar to the case of no electric field. However, as the strength of the electric field is enhanced, three distinctive regions of drag coefficients are found to exist. For a vertical (z-direction) electric field, the three particles separate out due to electrostatic repulsion and their drag increases as the inverse of the Mason number increase. Since the particles move far away from each other, the drag coefficient tends towards unity. In contrary, with horizontal (x-direction) electric field, the particles tend to come close and chain, and thus, the drag coefficient decreases with increase in the inverse of the Mason number. With even higher $Mn^{-1}$, the particles have already chained at the position $(z=10)$ where the drag coefficient is evaluated, thereby, maintaining a constant drag. For an XZ-directed electric field, a similar trend is observed and may be explained with similar reasoning as in the case of the horizontally applied electric field. However, the former case shows a further decrease in the drag coefficient as compared to the latter for very high values of $Mn^{-1}$. This is attributed to the inclined alignment of the middle particle (as seen in figure 6b) which leads to a net decrease in its projected area, thereby causing lesser interaction with the impeding fluid.

Figure 7b denotes the net volume flow rate due to the particle sedimenting motion and applied electric field orientation for an asymmetrically-arranged 3-particle system at a particular vertical height $(z=50)$. This volume flow rate is calculated by considering a box with a rectangular grid in the vicinity of the instantaneous particle positions at $(z=50)$ and averaging out the x-components of the fluid domain velocities, evaluated at all the grid points, multiplied by half the cell size. The fact that different electric field orientations at lower $Mn^{-1}$ have no effect on the flow rate is intuitive and has also been observed in figure 6a for drag coefficient evaluation. However, as the strength of the electric field is enhanced, we find that the volume flow rate varies drastically with field directions. It must be appreciated that the flow rate is established due to the initial asymmetrical arrangements of the particles. For a horizontal field, the particles are attracted closer to each other, and thus, the flow rate initially decreases. However, at very high field strength, the effect of the asymmetrical positioning of the particles is enhanced since even the left away particle is now strongly attracted through the surrounding fluid, resulting in an enhanced flow rate. A similar trend is observed for a vertical electric field where the repulsive forces among the particles replace the attractive forces in the previous case. Besides, slightly higher flow rate is observed in this case since the particles rotate and rearrange them along the vertical direction, thereby, displacing a higher quantity of the surrounding fluid. However, for an inclined field, the volume flow rate remains more or less similar to the case without the electric field (signified by a less $Mn^{-1}$ value). This is because, even with the presence of the field, the particles tend to arrange (along the inclined field) nearly in the same fashion as that without the presence of the field. With a higher field though, the rearrangement rate is faster inducing a higher flow rate. It must be noted that the volume flow rate of the surrounding fluid is merely due to the net displacement of the initial center of mass of the particle system in the horizontal direction, which is attributed to the particle rotation and hydrodynamic many-body interactions, even in absence of any net force along the horizontal direction. Both the volume flow rate and the drag coefficient would vary along the sedimenting height. We have selected a particular height where the effects are prominent while the qualitative reasoning for the variation of drag and volume flow rate remains consistent for any other vertical position.

## 5. Conclusions

The present work details the three dimensional dynamical evolution of finite sedimenting dielectric particles through an infinite Newtonian fluid medium under the presence of an external electric field. The many-body hydrodynamic interaction is included using the resistance matrix formulation analysis in purview of the Stokesian Dynamics, while the electrodynamic many-body problem employed the effective capacitance matrix formulation. The coupled hydrodynamic and electrodynamic interactions influence the particle trajectory and reveal intriguing chain formation that holds the basis for electrorheological effect. We have further shown the effect of an oblique electric field, an imposed shear and circular flow on the particle trajectories. This is a fundamental work which finds its significance in studies related to various fields like electrorheological and magnetorheological fluids, infinite suspensions and porous medium.

# Appendix A. Grand Capacitance Matrix Formulation

*Grand Potential Matrix formulation:* (Bonnecaze & Brady 1990)

$$\begin{pmatrix} \Phi_\alpha \\ \Phi_\beta \\ \cdot \\ \cdot \\ -G_\alpha^E \\ -G_\beta^E \\ \cdot \\ \cdot \\ O^E \\ O^E \\ \cdot \\ \cdot \end{pmatrix} = \begin{pmatrix} \hat{M}_{\Phi q}^{\alpha\alpha} \hat{M}_{\Phi q}^{\alpha\beta} ... \hat{M}_{\Phi S}^{\alpha\alpha} \hat{M}_{\Phi S}^{\alpha\beta} ... \hat{M}_{\Phi Q}^{\alpha\alpha} \hat{M}_{\Phi Q}^{\alpha\beta} ... \\ \hat{M}_{\Phi q}^{\beta\alpha} \hat{M}_{\Phi q}^{\beta\beta} ... \hat{M}_{\Phi S}^{\beta\alpha} \hat{M}_{\Phi S}^{\beta\beta} ... \hat{M}_{\Phi Q}^{\beta\alpha} \hat{M}_{\Phi Q}^{\beta\beta} ... \\ \cdot \quad \cdot \quad \cdot \quad \cdot \quad \cdot \quad \cdot \quad \cdot \\ \cdot \quad \cdot \quad \cdot \quad \cdot \quad \cdot \quad \cdot \quad \cdot \\ \hat{M}_{Gq}^{\alpha\alpha} \hat{M}_{Gq}^{\alpha\beta} ... \hat{M}_{GS}^{\alpha\alpha} \hat{M}_{GS}^{\alpha\beta} ... \hat{M}_{GQ}^{\alpha\alpha} \hat{M}_{GQ}^{\alpha\beta} ... \\ \hat{M}_{Gq}^{\beta\alpha} \hat{M}_{Gq}^{\beta\beta} ... \hat{M}_{GS}^{\beta\alpha} \hat{M}_{GS}^{\beta\beta} ... \hat{M}_{GQ}^{\beta\alpha} \hat{M}_{GQ}^{\beta\beta} ... \\ \cdot \quad \cdot \quad \cdot \quad \cdot \quad \cdot \quad \cdot \quad \cdot \\ \cdot \quad \cdot \quad \cdot \quad \cdot \quad \cdot \quad \cdot \quad \cdot \\ \hat{M}_{Oq}^{\alpha\alpha} \hat{M}_{Oq}^{\alpha\beta} ... \hat{M}_{OS}^{\alpha\alpha} \hat{M}_{OS}^{\alpha\beta} ... \hat{M}_{OQ}^{\alpha\alpha} \hat{M}_{OQ}^{\alpha\beta} ... \\ \hat{M}_{Oq}^{\beta\alpha} \hat{M}_{Oq}^{\beta\beta} ... \hat{M}_{OS}^{\beta\alpha} \hat{M}_{OS}^{\beta\beta} ... \hat{M}_{OQ}^{\beta\alpha} \hat{M}_{OQ}^{\beta\beta} ... \\ \cdot \quad \cdot \quad \cdot \quad \cdot \quad \cdot \quad \cdot \quad \cdot \\ \cdot \quad \cdot \quad \cdot \quad \cdot \quad \cdot \quad \cdot \quad \cdot \end{pmatrix} \cdot \begin{pmatrix} q_\alpha \\ q_\beta \\ \cdot \\ \cdot \\ S_\alpha \\ S_\beta \\ \cdot \\ \cdot \\ Q_\alpha \\ Q_\beta \\ \cdot \\ \cdot \end{pmatrix}$$

The grand potential matrix can be written in the following form depicting the nine sub-matrices:

$$\begin{pmatrix} \Phi \\ -G^E \\ O \end{pmatrix} = \begin{pmatrix} \hat{M}_{\Phi q} & \hat{M}_{\Phi S} & \hat{M}_{\Phi Q} \\ \hat{M}_{Gq} & \hat{M}_{GS} & \hat{M}_{GQ} \\ \hat{M}_{Oq} & \hat{M}_{OS} & \hat{M}_{OQ} \end{pmatrix} \cdot \begin{pmatrix} q \\ S \\ Q \end{pmatrix}$$

where $\Phi_\alpha$ represents the electrostatic potential at the centre of the particle $\alpha$ and $q_\alpha$ denotes it's charge. $G_\alpha^E$ represents the externally applied electric field evaluated at the particle center $\mathbf{x}_\alpha$, $O = -\nabla G^E$, $S_\alpha$ denotes it's dipole moment and $Q_\alpha$ is a measure of the particle quadrupole moment. The final reduced form of the potential matrix for the case of a constant electric field $O = 0$ is given by:

$$\begin{pmatrix} \Phi_\alpha \\ \Phi_\beta \\ \cdot \\ \cdot \\ -G^E_\alpha \\ -G^E_\beta \\ \cdot \\ \cdot \end{pmatrix} = \begin{pmatrix} \tilde{M}^{\alpha\alpha}_{\Phi q} \tilde{M}^{\alpha\beta}_{\Phi q} \ldots \tilde{M}^{\alpha\alpha}_{\Phi S} \tilde{M}^{\alpha\beta}_{\Phi S} \cdots \\ \tilde{M}^{\beta\alpha}_{\Phi q} \tilde{M}^{\beta\beta}_{\Phi q} \ldots \tilde{M}^{\beta\alpha}_{\Phi S} \tilde{M}^{\beta\beta}_{\Phi S} \cdots \\ \cdot \cdot \cdot \cdot \cdot \cdot \\ \cdot \cdot \cdot \cdot \cdot \cdot \\ \tilde{M}^{\alpha\alpha}_{Gq} \tilde{M}^{\alpha\beta}_{Gq} \ldots \tilde{M}^{\alpha\alpha}_{GS} \tilde{M}^{\alpha\beta}_{GS} \cdots \\ \tilde{M}^{\beta\alpha}_{Gq} \tilde{M}^{\beta\beta}_{Gq} \ldots \tilde{M}^{\beta\alpha}_{GS} \tilde{M}^{\beta\beta}_{GS} \cdots \\ \cdot \cdot \cdot \cdot \cdot \cdot \\ \cdot \cdot \cdot \cdot \cdot \cdot \end{pmatrix} \cdot \begin{pmatrix} q_\alpha \\ q_\beta \\ \cdot \\ \cdot \\ S_\alpha \\ S_\beta \\ \cdot \\ \cdot \end{pmatrix} \quad \Rightarrow \quad \begin{pmatrix} \Phi \\ -G^E \end{pmatrix} = \begin{pmatrix} \tilde{M}_{\Phi q} & \tilde{M}_{\Phi S} \\ \tilde{M}_{Gq} & \tilde{M}_{GS} \end{pmatrix} \cdot \begin{pmatrix} q \\ S \end{pmatrix}$$

where we have the reduced potential sub-matrices as follows:

$$\tilde{M}_{\Phi q} = \hat{M}_{\Phi q} - \hat{M}_{\Phi Q} \hat{M}^{-1}_{OQ} \hat{M}_{Oq}$$

$$\tilde{M}_{\Phi S} = \hat{M}_{\Phi S} - \hat{M}_{\Phi Q} \hat{M}^{-1}_{OQ} \hat{M}_{OS}$$

$$\tilde{M}_{Gq} = \hat{M}_{Gq} - \hat{M}_{GQ} \hat{M}^{-1}_{OQ} \hat{M}_{Oq}$$

$$\tilde{M}_{GS} = \hat{M}_{GS} - \hat{M}_{GQ} \hat{M}^{-1}_{OQ} \hat{M}_{OS}$$

The tensorial forms of the sub-matrices are shown below:

$$\hat{M}^{\alpha\beta}_{\Phi q} = 3\gamma^2 \left( \delta_{\alpha\beta} \left(1 + \frac{\lambda}{2\lambda_p}\right) + \left(1 - \delta_{\alpha\beta}\right) \frac{1}{r} \right)$$

$$_i\hat{M}^{\alpha\beta}_{\Phi S} = -\frac{3\gamma^2 e_i}{r^2}$$

$$_{ij}\hat{M}^{\alpha\beta}_{\Phi Q} = \frac{3\gamma^2}{2} \left( \frac{3e_i e_j}{r^3} - \frac{\delta_{ij}}{r^3} \right)$$

$$_i\hat{M}^{\alpha\beta}_{Gq} = \frac{3\gamma^2 e_i}{r^2}$$

$$_{ij}\hat{M}^{\alpha\beta}_{GS} = 3\gamma \delta_{ij} \delta_{\alpha\beta} - 3\gamma^2 \left(1 - \delta_{\alpha\beta}\right) \left( \frac{3e_i e_j}{r^3} - \frac{\delta_{ij}}{r^3} \right)$$

$$_{ijk}\hat{M}^{\alpha\beta}_{GQ} = 3\gamma^2 \left( \frac{15 e_i e_j e_k}{2r^4} - \frac{3}{2r^4} \left( e_i \delta_{jk} + e_j \delta_{ik} + e_k \delta_{ij} \right) \right)$$

$$_{ij}\hat{M}^{\alpha\beta}_{Oq} = 3\gamma^2 \left( \frac{3e_i e_j}{r^3} - \frac{\delta_{ij}}{r^3} \right)$$

$$_{ijk}\hat{M}^{\alpha\beta}_{OS} = 3\gamma^2 \left( \frac{3}{r^4} \left( e_i \delta_{jk} + e_j \delta_{ik} + e_k \delta_{ij} \right) - \frac{15 e_i e_j e_k}{r^4} \right)$$

$$_{ijkl}\hat{M}^{\alpha\beta}_{OQ} = 9\gamma \delta_{\alpha\beta} \delta_{ik} \delta_{jl} + \left(1 - \delta_{\alpha\beta}\right) \frac{3\gamma^2}{2} \left( \begin{array}{l} \frac{105 e_i e_j e_k e_l}{r^5} + \frac{3}{r^5} \left( \delta_{il}\delta_{jk} + \delta_{jl}\delta_{ik} + \delta_{kl}\delta_{ij} \right) \\ - \frac{15}{r^5} \left( e_i e_j \delta_{kl} + e_i e_k \delta_{jl} + e_j e_k \delta_{il} + e_i e_l \delta_{jk} + e_j e_l \delta_{ik} + e_k e_l \delta_{ij} \right) \end{array} \right)$$

where $\gamma = \frac{\lambda_p - \lambda}{\lambda_p + 2\lambda}$, is the Clausius-Mossotti (CM) factor, $\lambda_p$ and $\lambda$ being the dielectric constant/conductivity of the particle and the fluid respectively. In the present work, we have used the corresponding conductivity values for the particles and the fluid medium due to the fact that when a constant d.c. field is applied, the medium conductivities dominate the CM factor for the evaluation of the polarization forces (Davis 1992; Dhar et al. 2013). The distance between two particles labeled $\alpha$ and $\beta$ is nondimensionalized by the particle radius as in the hydrodynamic case and is represented by $r$. Similarly, the components of the unit vector joining the center of any particle to the reference particle is denoted by $e_i$. The final capacitance matrix $\mathfrak{C}$ is obtained by inverting the reduced form of the potential matrix as formulated above. There is no need to include near field electrostatic effects in the current simulation as the particles are considered to be dielectric materials. Near field singularity exists in the case of perfectly conducting particles.

# Appendix B. Grand Resistance Matrix Formulation

*Mobility Matrix Formulation:* (Durlofsky et al. 1987)

$$\begin{pmatrix} \mathbf{U}_\alpha - \mathbf{u}^\infty(\mathbf{x}^\alpha) \\ \mathbf{U}_\beta - \mathbf{u}^\infty(\mathbf{x}^\beta) \\ \cdot \\ \cdot \\ \Omega_\alpha - \Omega^\infty \\ \Omega_\beta - \Omega^\infty \\ \cdot \\ \cdot \\ -\mathbf{E}^\infty \\ -\mathbf{E}^\infty \\ \cdot \\ \cdot \end{pmatrix} = \begin{pmatrix} \mathbf{a}_{\alpha\alpha} \mathbf{a}_{\alpha\beta} \ldots \tilde{\mathbf{b}}_{\alpha\alpha} \tilde{\mathbf{b}}_{\alpha\beta} \ldots \tilde{\mathbf{g}}_{\alpha\alpha} \tilde{\mathbf{g}}_{\alpha\beta} \cdots \\ \mathbf{a}_{\beta\alpha} \mathbf{a}_{\beta\beta} \ldots \tilde{\mathbf{b}}_{\beta\alpha} \tilde{\mathbf{b}}_{\beta\beta} \ldots \tilde{\mathbf{g}}_{\beta\alpha} \tilde{\mathbf{g}}_{\beta\beta} \cdots \\ \cdot \cdot \cdot \cdot \cdot \cdot \\ \cdot \cdot \cdot \cdot \cdot \cdot \\ \mathbf{b}_{\alpha\alpha} \mathbf{b}_{\alpha\beta} \ldots \mathbf{c}_{\alpha\alpha} \mathbf{c}_{\alpha\beta} \ldots \tilde{\mathbf{h}}_{\alpha\alpha} \tilde{\mathbf{h}}_{\alpha\beta} \cdots \\ \mathbf{b}_{\beta\alpha} \mathbf{b}_{\beta\beta} \ldots \mathbf{c}_{\beta\alpha} \mathbf{c}_{\beta\beta} \ldots \tilde{\mathbf{h}}_{\beta\alpha} \tilde{\mathbf{h}}_{\beta\beta} \cdots \\ \cdot \cdot \cdot \cdot \cdot \cdot \\ \cdot \cdot \cdot \cdot \cdot \cdot \\ \mathbf{g}_{\alpha\alpha} \mathbf{g}_{\alpha\beta} \ldots \mathbf{h}_{\alpha\alpha} \mathbf{h}_{\alpha\beta} \ldots \mathbf{m}_{\alpha\alpha} \mathbf{m}_{\alpha\beta} \cdots \\ \mathbf{g}_{\beta\alpha} \mathbf{g}_{\beta\beta} \ldots \mathbf{h}_{\beta\alpha} \mathbf{h}_{\beta\beta} \ldots \mathbf{m}_{\beta\alpha} \mathbf{m}_{\beta\beta} \cdots \\ \cdot \cdot \cdot \cdot \cdot \cdot \\ \cdot \cdot \cdot \cdot \cdot \cdot \end{pmatrix} \cdot \begin{pmatrix} \mathbf{F}_\alpha \\ \mathbf{F}_\beta \\ \cdot \\ \cdot \\ \mathbf{L}_\alpha \\ \mathbf{L}_\beta \\ \cdot \\ \cdot \\ \mathbf{S}_\alpha \\ \mathbf{S}_\beta \\ \cdot \\ \cdot \end{pmatrix}$$

where $\mathbf{U}_\alpha - \mathbf{u}^\infty(\mathbf{x}_\alpha) = \begin{pmatrix} U_\alpha^1 - u_1^\infty(x_\alpha) \\ U_\alpha^2 - u_2^\infty(x_\alpha) \\ U_\alpha^3 - u_3^\infty(x_\alpha) \end{pmatrix}$, $\Omega_\alpha - \Omega^\infty = \begin{pmatrix} \Omega_\alpha^1 - \Omega_1^\infty \\ \Omega_\alpha^2 - \Omega_2^\infty \\ \Omega_\alpha^3 - \Omega_3^\infty \end{pmatrix}$, while the symmetric and traceless rate of strain tensor is written as $\mathbf{E}^\infty = \left(E_{11}^\infty - E_{33}^\infty, 2E_{12}^\infty, 2E_{13}^\infty, 2E_{23}^\infty, E_{22}^\infty - E_{33}^\infty\right)^T$, and $\mathbf{x}_\alpha = (x_\alpha^1, x_\alpha^2, x_\alpha^3)^T \ \forall \alpha \in [1, N]$ denotes the position vector of the centre of the $\alpha$ particle. Similarly, we have $\mathbf{F}_\alpha = \begin{pmatrix} F_\alpha^1 \\ F_\alpha^2 \\ F_\alpha^3 \end{pmatrix}$, $\mathbf{L}_\alpha = \begin{pmatrix} L_\alpha^1 \\ L_\alpha^2 \\ L_\alpha^3 \end{pmatrix}$, and $\mathbf{S}_\alpha = \left(S_\alpha^{11}, S_\alpha^{12}, S_\alpha^{13}, S_\alpha^{23}, S_\alpha^{22}\right)^T$, $\forall \alpha \in [1, N]$. Here, **a**, **b**, **c** represent $(3X3)$ matrices, **g**, **h** represent $(5X3)$ matrices and **m** denotes a $(5X5)$ matrix for each pair of $\alpha$ & $\beta$ particles. The matrices $\tilde{\mathbf{b}}, \tilde{\mathbf{g}}, \tilde{\mathbf{h}}$ are directly related to **b**, **g**, **h** respectively. The relations in terms of the (i,j)th element are shown below:

$\tilde{b}_{\alpha\beta}^{ij} = b_{\beta\alpha}^{ji}, \ \tilde{g}_{\alpha\beta}^{ij} = g_{\beta\alpha}^{ji}, \ \tilde{h}_{\alpha\beta}^{ij} = h_{\beta\alpha}^{ji}$.

Given below are the tensorial forms of the calculated mobility matrix components. The distance between a pair of particles has been non-dimensionalized by the particle radius and is represented by $r$. The components of the unit vector joining any particle centre to the centre of the reference particle is denoted by $e_i$:

$$a_{ij}^{\alpha\beta} = x_{\alpha\beta}^a e_i e_j + y_{\alpha\beta}^a \left(\delta_{ij} - e_i e_j\right)$$

$$b_{ij}^{\alpha\beta} = y_{\alpha\beta}^b \varepsilon_{ijk} e_k$$

$$c_{ij}^{\alpha\beta} = x_{\alpha\beta}^c e_i e_j + y_{\alpha\beta}^c \left(\delta_{ij} - e_i e_j\right)$$

$$g_{ijk}^{\alpha\beta} = x_{\alpha\beta}^g \left(e_i e_j - \frac{1}{3}\delta_{ij}\right) e_k + y_{\alpha\beta}^g \left(e_i \delta_{jk} + e_j \delta_{ik} - 2 e_i e_j e_k\right)$$

$$h_{ijk}^{\alpha\beta} = y_{\alpha\beta}^h \left(e_i \varepsilon_{jkl} e_l + e_j \varepsilon_{ikl} e_l\right)$$

$$m_{ijkl}^{\alpha\beta} = \frac{3}{2} x_{\alpha\beta}^m \left(e_i e_j - \frac{1}{3}\delta_{ij}\right)\left(e_k e_l - \frac{1}{3}\delta_{kl}\right) + \frac{1}{2} y_{\alpha\beta}^m \left(e_i \delta_{jl} e_k + e_j \delta_{il} e_k + e_i \delta_{jk} e_l + e_j \delta_{ik} e_l - 4 e_i e_j e_k e_l\right)$$

$$+ \frac{1}{2} z_{\alpha\beta}^m \left(\delta_{ik}\delta_{jl} + \delta_{jk}\delta_{il} - \delta_{ij}\delta_{kl} + e_i e_j \delta_{kl} + \delta_{ij} e_k e_l + e_i e_j e_k e_l - e_i \delta_{jl} e_k - e_j \delta_{il} e_k - e_i \delta_{jk} e_l - e_j \delta_{ik} e_l\right)$$

The multiplication factors for each type of matrices are given by:

$$x_{11}^a = x_{22}^a = 1, \quad x_{12}^a = x_{21}^a = \frac{3}{2}r^{-1} - r^{-3},$$

$$y_{11}^a = y_{22}^a = 1, \quad y_{12}^a = y_{21}^a = \frac{3}{4}r^{-1} + \frac{1}{2}r^{-3},$$

$$y_{11}^b = -y_{22}^b = 0, \quad y_{12}^b = -y_{21}^b = -\frac{3}{4}r^{-2},$$

$$x_{11}^c = x_{22}^c = \frac{3}{4}, \quad x_{12}^c = x_{21}^c = \frac{3}{4}r^{-3},$$

$$y_{11}^c = y_{22}^c = \frac{3}{4}, \quad y_{12}^c = y_{21}^c = -\frac{3}{8}r^{-3},$$

$$x_{11}^g = -x_{22}^g = 0, \quad x_{12}^g = -x_{21}^g = \frac{9}{4}r^{-2} - \frac{18}{5}r^{-4},$$

$$y_{11}^g = -y_{22}^g = 0, \quad y_{12}^g = -y_{21}^g = \frac{6}{5}r^{-4},$$

$$y_{11}^h = y_{22}^h = 0, \quad y_{12}^h = y_{21}^h = -\frac{9}{8}r^{-3},$$

$$x_{11}^m = x_{22}^m = \frac{9}{10}, \quad x_{12}^m = x_{21}^m = -\frac{9}{2}r^{-3} + \frac{54}{5}r^{-5},$$

$$y_{11}^m = y_{22}^m = \frac{9}{10}, \quad y_{12}^m = y_{21}^m = \frac{9}{4}r^{-3} - \frac{36}{5}r^{-5},$$

$$z_{11}^m = z_{22}^m = \frac{9}{10}, \quad z_{12}^m = z_{21}^m = \frac{9}{5}r^{-5}.$$

The Lubrication correction is made to the grand resistance matrix by adding to it the difference between the analytically obtained exact resistance matrix and the inverse of the far field mobility matrix, both calculated for the case of two spheres near contact, in a pairwise manner.

$$\mathbf{R}_{grand} = (\mathbf{M}^\infty)^{-1}$$
$$\mathbf{R}_{exact} = \mathbf{R}_{grand} + (R_{2B}^{exact} - R_{2B}^{far-field})$$

## Appendix C: Velocity Field Distribution

We present here the expressions for calculating the velocity field in the fluid domain. The fluid velocity at any general point $\mathbf{x}$ in the fluid medium due to the immersed particles is given as:

$$u_i(\mathbf{x}) = u_i^\infty(\mathbf{x}) + \frac{1}{8\pi\mu}\left(\sum_{\alpha=1}^{N}\left(1+\frac{1}{6}a^2\nabla^2\right)J_{ij}F_j^\alpha + R_{ij}L_j^\alpha + \left(1+\frac{1}{10}a^2\nabla^2\right)K_{ijk}S_{jk}^\alpha + ...\right)$$

where $u_i^\infty(\mathbf{x})$ is the $i^{th}$ component of the imposed fluid flow velocity at the point $\mathbf{x}$, and the higher order tensors and their derivatives are given by:

$$R_{ij} = \varepsilon_{ijk}\frac{e_k}{r^2}$$

$$J_{ij} = \frac{1}{r}\left(\delta_{ij} + e_i e_j\right)$$

$$K_{ijk} = \frac{1}{r^2}\left(e_i\delta_{jk} - 3e_i e_j e_k\right)$$

$$\nabla^2 J_{ij} = \frac{1}{r^3}\left(\delta_{ij} - 3e_i e_j\right)$$

$$\nabla^2 K_{ijk} = -\frac{6}{r^4}(e_k\delta_{ij} + e_j\delta_{ik})$$

Here $r$ is the distance of the point $\mathbf{x}$ from the center of particle $\alpha$ and the summation in the equation indicates a sum over all the $N$ particles in the fluid media. The unit vector joining the point and a particle center is denoted by $e$.

With the known forces, torques and stresslets on the particles, the fluid velocity at any point can be calculated directly using the above stated equation.